\documentclass[reprint, aps,pra,twocolumn,superscriptaddress]{revtex4-1}
\usepackage{braket}
\usepackage{amssymb}
\usepackage{amsmath}
\usepackage{epsfig}
\usepackage{color}
\usepackage{graphics, graphicx}
\usepackage{bbold}
\usepackage{psfrag}
\usepackage{mathcomp}
\usepackage{subfigure}
\usepackage{verbatim}
\usepackage[colorlinks, citecolor=blue]{hyperref}
\usepackage[normalem]{ulem}
\usepackage{bm}
\def\cp#1{\mathbf{#1}}

\begin{document}

\title{Chiral Quantum Droplet in a Spin-Orbit Coupled Bose Gas}
\author{Tianqi Luo}
\affiliation{Beijing National Laboratory for Condensed Matter Physics, Institute of Physics, Chinese Academy of Sciences, Beijing 100190, China}
\affiliation{School of Physical Sciences, University of Chinese Academy of Sciences, Beijing 100049, China}
\author{Xiaoling Cui}
\email{xlcui@iphy.ac.cn} 
\affiliation{Beijing National Laboratory for Condensed Matter Physics, Institute of Physics, Chinese Academy of Sciences, Beijing 100190,  China}
\date{\today}

\begin{abstract}
We report the formation of {\it chiral} quantum droplet in a spin-orbit coupled Bose gas, where the system turns to a self-bound droplet when moving towards a particular direction and remains gaseous otherwise.  The chirality arises from the breaking of Galilean invariance  by spin-orbit coupling, which enables the system to dynamically adjust its condensation momentum and spin polarization in response to its velocity.
As a result, only towards a specific moving direction and beyond a critical velocity, the acquired spin polarization can trigger collective interactions sufficient for self-binding and drive a first-order transition from gas to droplet. We have mapped out a phase diagram of droplet, gas and their coexistence for realistic spin-orbit coupled $^{39}$K mixtures with tunable moving velocity and magnetic detuning. Our results have revealed the emergence of chirality in spin-orbit coupled quantum gases, which shed light on general chiral phenomena in moving systems with  broken Galilean invariance.    
\end{abstract}
\maketitle

\section{Introduction}

Quantum droplets, a novel form of self-bound quantum matter, have emerged as a paradigm for exploring beyond-mean-field physics in ultracold atomic gases. Such droplets are stabilized by a critical balance between mean-field attraction and repulsive quantum fluctuations, as predicted by Petrov’s seminal theory\cite{Petrov}, and have been successfully realized in magnetic dipolar bosons\cite{Pfau_PRL2016, Pfau_Nature2016, Ferlaino_PRX2016, Modugno_PRL2019, Pfau_PRX2019, Ferlaino_PRX2019} and alkali Bose-Bose mixtures\cite{Tarruell_Science2018,Tarruell_PRL2018,Fattori_PRL2018,Fort_PRR2019, Fort_CM2020, Wang_PRR2021}. 
The formation of these droplets in the dilute regime is facilitated by the existence of spin degrees of freedom or multiple interaction channels, such that the Lee-Huang-Yang (LHY) correction from quantum fluctuations is comparable with the mean-field strength to reach a balance for droplet stabilization\cite{Petrov}. Moreover, the spin degrees of freedom has also been shown to induce intriguing quantum  phenomena such as liquid-gas coexistence\cite{Gu,He, Giorgini},  exotic spin textures\cite{Bisset, Smith}, Borromean binding\cite{Ma_1}, mixed-bubble\cite{Naidon} and core-shell\cite{Ma_2} structures, etc. Spin-orbit coupling (SOC)\cite{soc_review}, which intertwines a system’s spin and motion, can even amplify the role of spin and is expected to  intrigue novel self-bound phases with nontrivial spin-motion correlations. Despite a number of existing studies on quantum droplets with SOC\cite{Li, Cui, Reimann, Mazzanti, Yin, Zeng, Mishra}, the dynamical consequences of SOC -- particularly under motion -- have been rarely explored.

A key feature of SOC systems is the broken Galilean invariance\cite{Wu,Zheng}, which can give rise to velocity-dependent phenomena that are absent in conventional quantum fluids. For instance, a recent experiment on 1D spin-orbit coupled Bose gases has observed chiral solitons that emerge exclusively when the system is kicked towards one direction, which is attributed to the SOC-induced current-density interactions\cite{Frolian}. Similarly, chiral Raman coupling for SOC has been also realized experimentally, where the SOC field sensitively depends on magnetic field orientation\cite{Zhang}.  These advances highlight the intrinsic topological nature of SOC systems with directional selectivity under motion.

In this work, we report a topological self-bound state, i.e., the chiral quantum droplet,  in spin-orbit coupled Bose-Bose mixtures, where the droplet formation occurs only if the system moves towards a particular direction. Such directional asymmetry arises from the breaking of Galilean invariance under SOC, such that a moving system can have very different intrinsic properties from the static one. As a result, the system can dynamically adjust its condensation momentum and spin polarization in response to a finite moving velocity. 
Only by moving towards a particular direction and beyond a critical velocity, the system can acquire a proper spin polarization that triggers sufficient mean-field attraction to support self-binding. In this way, the velocity naturally drives a first-order transition from gaseous to droplet states. Taking the realistic  spin-orbit-coupled $^{39}$K Bose-Bose mixtures for example, we have mapped out the phase diagram including droplet, gas and their coexistent phases by tuning the moving velocity and magnetic detuning. Our results not only establish a pathway to observe chiral droplets in current experiments, but also suggest a broader paradigm to engineer topological quantum matter in moving systems with broken Galilean invariance.

The rest part of the paper is organized as follows. In Section \ref{sec_model} we present the theoretical model and formula for spin-orbit coupled bosons in a moving frame. Results of chiral quantum droplet and the associated phase diagram are provided in Section \ref{sec_chiral}. Finally we make further remarks in Section \ref{sec_summary}. 

\section{Model and Formula} \label{sec_model}

We consider the two-species ($\uparrow,\downarrow$) bosons with spin-orbit coupling (SOC) along $x$ direction and also moving along this direction (note that a moving system along $y$ or $z$ direction is Galilean invariant and thus equivalent to that  without moving).  As in a recent experiment\cite{Frolian}, a finite moving velocity ($v$) can be created by applying two additional Raman lasers counter-propagating along $x$, which impart a finite center-of-mass momentum ($k_x=mv$) to the system.  In the moving frame, the Hamiltonian can be written as  (taking $\hbar=1$):
\begin{eqnarray}
H&=&H_0+U,\nonumber\\
H_0&=& \int d{\cp r}\left\{ \sum_{\sigma}\psi_{\sigma}^\dagger({\cp r}) \left( \frac{({\cp p}+q\sigma_z\hat{x})^2}{2m} -v p_x\right) \psi_{\sigma}({\cp r}) \right. \nonumber\\
&&\ \ \ \left. -\sum_{\sigma\sigma'}\psi_{\sigma}^\dagger({\cp r}) (\Omega\sigma_x+\delta \sigma_z)_{\sigma\sigma'} \psi_{\sigma'}({\cp r})\right\} ;\nonumber\\
U&=&\frac{1}{2} \int d{\cp r} \sum_{\sigma\sigma'}g_{\sigma\sigma'}\psi_{\sigma}^\dagger({\cp r})\psi_{\sigma'}^\dagger ({\cp r})\psi_{\sigma'}({\cp r})\psi_{\sigma} ({\cp r}). \label{H}
\end{eqnarray}
where $ \psi_{\sigma}^\dagger(\mathbf{r}) $ is the creation operator of spin-$\sigma$, and $\sigma_{i}$ ($i=x,y,z$) are Pauli matrices; ${\cp p}$ is the momentum operator and $p_x$ is its $x$-component; $q$ and $\Omega$ are respectively the transferred momentum and strength of SOC field; $\delta$ is the magnetic detuning; $g_{\sigma\sigma'}=4\pi a_{\sigma\sigma'}/m$ is the coupling strength between $\sigma$ and $\sigma'$ species, with $a_{\sigma\sigma'}$ the according scattering length. 

For the single-particle part, we will consider the single minimum (or plane-wave) case of dispersion relation, to facilitate later discussions of quantum fluctuations. For the interaction part, we will consider $g_{\uparrow\uparrow},g_{\downarrow\downarrow}>0$ and $g_{\uparrow\downarrow}<0$, such that $\Delta g\equiv g_{\uparrow\downarrow}+\sqrt{g_{\uparrow\uparrow}g_{\downarrow\downarrow}}<0$ and a quantum droplet can be supported in 3D in the absence of SOC\cite{Petrov}. Specifically, we take the two hyperfine states of $^{39}$K atoms, $|\uparrow\rangle\equiv|F=1,m_F=-1\rangle,\ |\downarrow\rangle\equiv|F=1,m_F=0\rangle$, as studied in quantum droplet experiments\cite{Tarruell_Science2018,Tarruell_PRL2018,Fattori_PRL2018}. In this case,  $a_{\uparrow\uparrow}=35a_B,\ a_{\uparrow\downarrow}=-53a_B$ ($a_B$ is the Bohr radius), and $a_{\downarrow\downarrow}$ is tunable by magnetic field near $B\sim57$G. For SOC, we consider  two counter-propagating Raman beams with wavelength $\lambda=769$nm\cite{Frolian} and opposite momenta $q$ and $-q$ $(q=2\pi/\lambda)$, giving the recoil energy $E_q=q^2/(2m)$.

For a system in the thermodynamic limit, its property  depends on three variables: total density $n\equiv n_{\uparrow}+n_{\downarrow}$, spin polarization $S\equiv(n_{\uparrow}-n_{\downarrow})/n$ and condensation momentum ${\cp p_0}\equiv(p_{0x},0,0)$. At a given $\{n, S, p_{0x} \}$,  the total energy density $\epsilon\equiv E/V$ are composed by the mean-field part and LHY correction from quantum flunctuations:
\begin{equation}
\epsilon=\epsilon_{\rm mf}+\epsilon_{\rm LHY}.
\end{equation}
For the mean-field part, we have
\begin{equation}
\begin{split}
\epsilon_{\rm mf}&=\frac{(p_{0x}+q)^2}{2m}n_\uparrow+\frac{(p_{0x}-q)^2}{2m}n_\downarrow-2\Omega\sqrt{n_\uparrow n_\downarrow}-\delta(n_\uparrow-n_\downarrow)\\
&\ -vp_{0x}n+\frac{1}{2}(g_{\uparrow\uparrow}n_\uparrow^2+g_{\downarrow\downarrow}n_\downarrow^2+2g_{\uparrow\downarrow}n_\uparrow n_\downarrow)\\
&=n\left[\frac{p_{0x}^2+q^2}{2m}-\sqrt{1-S^2}\Omega+(\frac{p_{0x}q}{m}-\delta)S-vp_{0x}\right]\\
&\ +\frac{g_0n^2}{2}\left[(S-\beta)^2+\frac{g_{\uparrow\uparrow}g_{\downarrow\downarrow}-g_{\uparrow\downarrow}^2}{4g_0^2}\right]
\label{e_mf}
\end{split}
\end{equation}
with $g_0=(g_{\uparrow\uparrow}+g_{\downarrow\downarrow}-2g_{\uparrow\downarrow})/4$, $\beta=(g_{\downarrow\downarrow}-g_{\uparrow\uparrow})/4g_0$. In the absence of SOC and magnetic detuning, the minimization of $\epsilon_{\rm mf}$ gives rise to $S=\beta$, and thus the interaction part is associated with a negative effective coupling $(g_{\uparrow\uparrow}g_{\downarrow\downarrow}-g_{\uparrow\downarrow}^2)/(4g_0)$, a precursor for droplet formation\cite{Petrov}.   However, when turn on SOC and drive the system with a finite moving velocity,  $S$ can be highly tunable and the droplet formation can become chiral, as we will discuss in Section \ref{sec_chiral}.


\begin{widetext}
The LHY correction to mean-field energy can be obtained via standard Bogoliubov analysis, which has been derived for both plane-wave\cite{Zheng, Torma} and stripe\cite{Mazzanti, Yin} phases of spin-orbit coupled bosons. For the present plane-wave system in a moving frame,  the LHY energy density follows
\begin{equation}
\begin{split}
\epsilon_{\rm LHY}&=\frac{1}{V}\sum_{k_x>0;k_y,k_z} \left(\sum_{\sigma, \sigma'} \frac{g^2_{\sigma \sigma'}n_\sigma n_\sigma'}{2\epsilon_{\cp k}^{(0)}}-\Big(2\epsilon_{\cp k}^{(0)}-\frac{2k_xp_{0x}}{m}+2vk_x+\Omega \frac{(n_\uparrow+n_\downarrow)}{\sqrt{n_\uparrow  n_\downarrow}}+g_{\uparrow\uparrow}n_\uparrow+g_{\downarrow\downarrow}n_\downarrow\Big)+E_+(-{\cp k})+E_-(-{\cp k})\right)
\end{split}
\end{equation}
Here $\epsilon_{\cp k}^{(0)}={\cp k}^2/(2m)$, and the excitation spectra can be obtained from the eigen-values ($\{E_+({\cp k}), -E_+(-{\cp k}), E_-({\cp k}), -E_-(-{\cp k})\}$) of the following $4\times 4$ matrix:
\begin{equation}
{\cal M} = \begin{bmatrix}
\epsilon_{\cp k}^{(0)}+h_{\uparrow}(k_x,p_{0x}+q) & g_{\uparrow\uparrow}n_\uparrow & g_{\uparrow\downarrow}\sqrt{n_\uparrow n_\downarrow}-\Omega & g_{\uparrow\downarrow}\sqrt{n_\uparrow n_\downarrow} \\
-g_{\uparrow\uparrow}n_\uparrow & -\epsilon_{\cp k}^{(0)}-h_{\uparrow}(-k_x,p_{0x}+q) & -g_{\uparrow\downarrow}\sqrt{n_\uparrow n_\downarrow} & -g_{\uparrow\downarrow}\sqrt{n_\uparrow n_\downarrow}+\Omega \\
g_{\uparrow\downarrow}\sqrt{n_\uparrow n_\downarrow}-\Omega & g_{\uparrow\downarrow}\sqrt{n_\uparrow n_\downarrow} & \epsilon_{\cp k}^{(0)}+h_{\downarrow}(k_x,p_{0x}-q) & g_{\downarrow\downarrow}n_\downarrow \\
-g_{\uparrow\downarrow}\sqrt{n_\uparrow n_\downarrow} & -g_{\uparrow\downarrow}\sqrt{n_\uparrow n_\downarrow}+\Omega & -g_{\downarrow\downarrow}n_\downarrow & -\epsilon_{\cp k}^{(0)}-h_{\downarrow}(-k_x,p_{0x}-q)
\end{bmatrix}
\label{eq:HBog}
\end{equation}
where
\begin{equation}
h_{\sigma}(k_1,k_2)=\frac{k_1k_2}{m}+\Omega\sqrt\frac{n_{\bar{\sigma}}}{n_\sigma}- vk_1+g_{\sigma\sigma}n_\sigma 
\label{eq:(2))}
\end{equation}
The ground state of the system is determined by minimizing $\epsilon$ in terms of three variables $\{n, S, p_{0x}\}$. 


In the low density limit ($n\rightarrow 0$), the stability of gaseous state relies on an effective interaction:
\begin{equation}
g_{\rm eff}\equiv \frac{d^2\epsilon}{d n^2}|_{n\rightarrow 0} = g_{\rm eff}^{\rm mf}+g_{\rm eff}^{\rm LHY}, \label{geff}
\end{equation} 
where  $g_{\rm eff}^{\rm mf}$ and $g_{\rm eff}^{\rm LHY}$ are respectively the mean-field and LHY contributions. The mean-field effective interaction can be written as  
\begin{equation} 
g_{\rm eff}^{\rm mf}=\sum_{\sigma\sigma'=\{\uparrow,\downarrow\}} c_{\sigma\sigma'} g_{\sigma\sigma'}, \label{g_mf}
\end{equation}
where the coefficients $c_{\sigma\sigma'}$ are related to the spin polarization as $c_{\uparrow\uparrow}=(S+1)^2/4$, $c_{\downarrow\downarrow}=(1-S)^2/4$ and $c_{\uparrow\downarrow}=c_{\downarrow\uparrow}=(1-S^2)/4$. 

The LHY contribution, $g_{\rm eff}^{\rm LHY}$, can be obtained via second-order perturbation theory. Namely, we start from two bosons at the single-particle minimum (${\cp p}_0=(p_{0x},0,0)$) in the lowest helicity branch, and consider their virtual excitations to other momentum states in both branches. Specifically, the virtual scattering is described by the Hamiltonian:
\begin{eqnarray}
H^{(0)}&=&\frac{1}{2V}\sum_{\alpha\beta=\{\pm\}}\sum_{{\cp k}}g_{\alpha\beta}(k_x)\left(\phi^{\dag}_{\alpha;{\cp p_0}+{\cp k}}\phi^{\dag}_{\beta;{\cp p_0}-{\cp k}} \phi_{-;{\cp p_0}}\phi_{-;{\cp p_0}} +h.c. \right)\nonumber\\
&=&\frac{1}{2V}\sum_{\alpha\beta=\{\pm\}}\sum_{k_x>0, k_y, k_z}\tilde{g}_{\alpha\beta}(k_x)\left(\phi^{\dag}_{\alpha;{\cp p_0}+{\cp k}}\phi^{\dag}_{\beta;{\cp p_0}-{\cp k}} \phi_{-;{\cp p_0}}\phi_{-;{\cp p_0}} +h.c.\right)
\end{eqnarray}
where $\phi^{\dag}_{\alpha;{\cp k}}$ is to create a single-particle eigenstate at helicity branch $\alpha$ and momentum ${\cp k}$, with eigen-energy 
\begin{equation}
\varepsilon_{\alpha}(k)=\frac{\bm k^2+q^2}{2m}-vk_x+\alpha\sqrt{\Omega^2+(\frac{qk_x}{m}-\delta)^2};
\label{eq:(4))}
\end{equation}
$g_{\alpha\beta}$ is the according coupling constant and $\tilde{g}_{\alpha\beta}(k_x)=g_{\alpha\beta}(k_x)+g_{\beta\alpha}(-k_x)$. 
This gives rise to a second-order energy correction $\epsilon_{\rm LHY}=g_{\rm eff}^{\rm LHY} n^2/2$, where 
\begin{equation} 
g_{\rm eff}^{\rm LHY}=\frac{1}{2V}\sum_{\alpha\beta=\{\pm\}}\sum_{k_x>0,k_y,k_z}  \frac{|\tilde{g}_{\alpha\beta}(k_x)|^2}{2\varepsilon_-(\bm p_0)-\varepsilon_{\alpha}(\bm p_0+\bm k)-\varepsilon_{\beta}(\bm p_0-\bm k)} +\frac{1}{V}\sum_{\sigma\sigma'=\{\uparrow,\downarrow\}}\sum_{\cp k}\frac{c_{\sigma\sigma'}g_{\sigma\sigma'}^2}{2\epsilon_{\cp k}^{(0)}} \label{g_LHY}
\end{equation}
Note that the ultraviolet divergence in the first term of above equation can be well eliminated by adding the second term, which is from the renormalization of  $g_{\rm eff}^{\rm mf}$ 
(see Eq.(\ref{g_mf})) in the mean-field energy via $g_{\sigma\sigma'}\rightarrow g_{\sigma\sigma'} + (1/V)\sum_{\cp k} g_{\sigma\sigma'}^2/(2\epsilon_{\cp k}^{(0)})$. We have checked numerically that $g_{\rm eff}^{\rm LHY}$ from Eq.(\ref{g_LHY}) is consistent with  the result from second-order derivative of  $\epsilon_{\rm LHY}$ to $n$ in $n\rightarrow 0$ limit.

\end{widetext}




\section{Results} \label{sec_chiral}

In this section, we present the result of chiral quantum droplet in a moving system of spin-orbit coupled bosons. To better understand its physical origin, we will start from the low-density limit and show how the moving velocity changes effective interaction (Eq.(\ref{geff})). Then we will present the full phase diagram based on numerical calculations for all density regimes.

\subsection{Low density limit}

\begin{figure}[htbp] 
    \centering 
    \includegraphics[width=9cm]{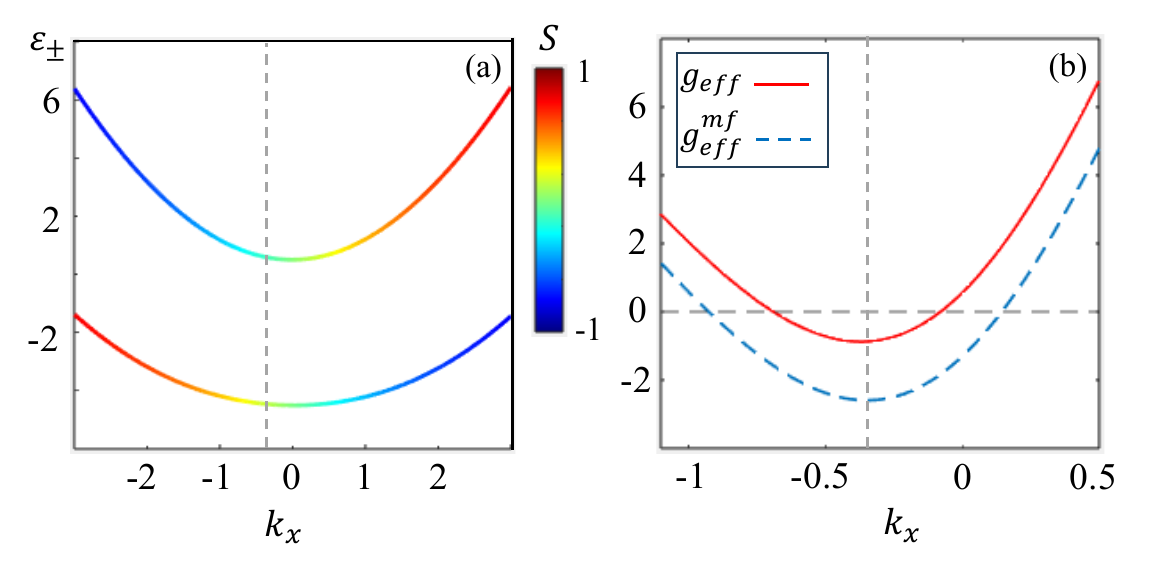} 
    \caption{(Color online.) (a)Single-particle dispersion along $k_x$ ($k_y=k_z=0$) for static system with SOC. 
According colors denote spin polarizations $S$. Here we take $\Omega=5E_q$ and $\delta=0.07E_q$.   (b) Total and mean-field effective interactions as functions of $k_x$. Here we take the same $\Omega$ and $\delta$ as in (a), and the tunable scattering length as   $a_{\downarrow\downarrow}=65a_B$. The units of energy, momentum, and effective interaction are $E_q$, $q$ and $4\pi a_B/m$, respectively.}  \label{fig_spectrum} 
\end{figure}

In the low-density limit, the system is dominated by single-particle physics. In Fig.\ref{fig_spectrum}(a), we plot out the typical single-particle energy spectrum $\varepsilon_{\pm}$ along $k_x$ and the spin polarization $S$ for static system ($v=0$). We can see that $S$ changes sensitively with $k_x$ for both helicity branches, due to the spin-momentum locking under SOC. For the lowest helicity branch, we have
\begin{equation}
S=\frac{\delta-qk_x/m}{\sqrt{\Omega^2+(\delta-qk_x/m)^2}}
\label{eq:(3))}
\end{equation}
Such highly tunable $S$ has also been revealed in two-species bosons under Rabi coupling and magnetic detuning fields, which can lead to  liquid-gas transition and coexistence at zero temperature\cite{Gu}. In comparison, here our focus is the chiral character due to the interplay of SOC and moving velocity, as discussed below.  

In combination with sensitive dependence of $g_{\rm eff}^{\rm mf}$ on $S$ (see Eq.(\ref{g_mf})), we  can see that $g_{\rm eff}^{\rm mf}$ closely relies on $k_x$, as shown in Fig.\ref{fig_spectrum}(b). In particular, $g_{\rm eff}^{\rm mf}$ can be tuned to negative near $S\sim \beta$, as marked by vertical dashed line in  Fig.\ref{fig_spectrum}(a,b). The inclusion of LHY contribution will uplift the effective interaction, as shown by solid curve in Fig.\ref{fig_spectrum}(b). Nevertheless, near the vertical dashed line the total $g_{\rm eff}$ is still negative, suggesting the droplet formation once the condensation momentum $p_{0x}$ is tuned nearby this location. In the following, we will show that $p_{0x}$ can be efficiently tuned by a finite moving velocity. 

\begin{figure}[htbp] 
    \centering 
    \includegraphics[width=9cm]{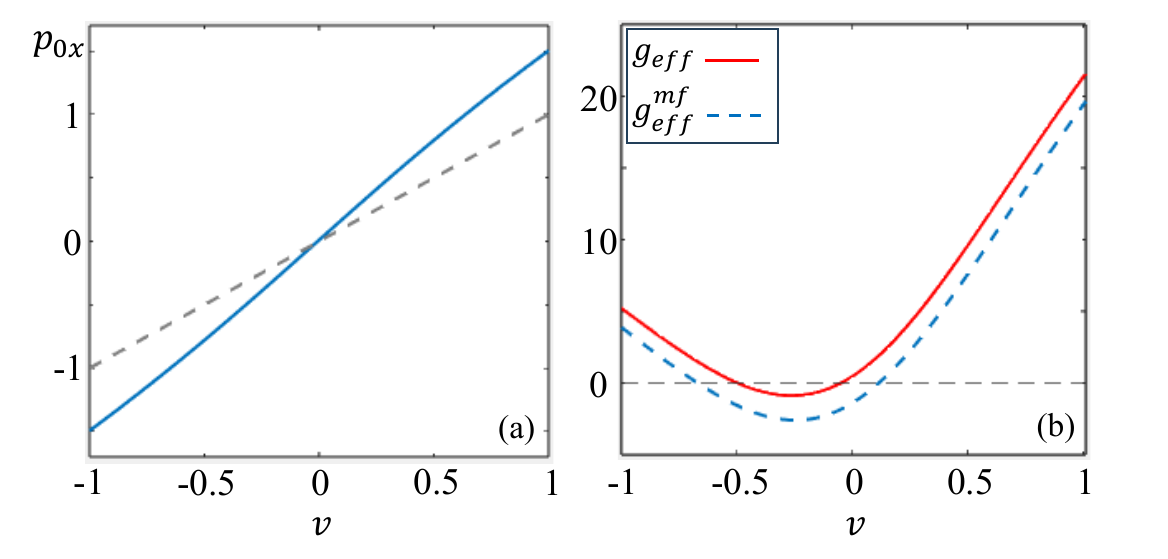} 
    \caption{(Color online.) (a) Condensation momentum  ($p_{0x}$) as a function of moving velocity $v$ in low-density limit. Solid and Dashed lines respectively show the true $p_{0x}$ and the linear dependence $p_{0x}=mv$. Their deviation comes from the fact that the system breaks  Galilean invariance under SOC. (b) Mean-field ($g^{\rm mf}_{\rm eff}$) and total ($g_{\rm eff}$) effective interactions  as functions of moving velocity ($v$) in low-density limit. The single-particle and interaction parameters are the same as in Fig.\ref{fig_spectrum}. The units of momentum, velocity and effective interaction are $q$, $q/m$ and $4\pi a_B/m$, respectively. 
    }   \label{fig_v} 
\end{figure}

Under a finite moving velocity ($v\neq0$), the system can dynamically adjust its condensation momentum ($p_{0x}$) in response to $v$. In the low density regime, $p_{0x}$ is determined by the minimization of single-particle energy $\varepsilon_-$. Its dependence on $v$ is shown by solid line in Fig.\ref{fig_v}(a). Due to the breaking of Galilean invariance, $p_{0x}$ deviates from the trivial linear dependence $p_{0x}=mv$ (dashed line in Fig.\ref{fig_v}(a)). Since $p_{0x}$ changes with $v$, $g_{\rm eff}$ can also be conveniently tuned by $v$, as shown in Fig.\ref{fig_v}(b). In this way,  the moving velocity offers an efficient tool to tune the effective interaction in low density limit. Around certain range of negative $v$, $g_{\rm eff}$ can be tuned to negative. This suggests the instability of gaseous state at small $n$ and therefore the tendency of droplet formation as ground state instead.

Here we remark that the velocity-dependences of condensation momentum and effective interaction are direct manifestations of Galilean invariance breaking, which tells that the $-vp_x$ term in the Hamiltonian (Eq.(\ref{H})) cannot be gauged away and thus a  moving system can have very different intrinsic properties from the static one. This is why the droplet formation in SOC system crucially depends on the moving velocity $v$. To be more specific, only for $v$ moving along $-x$ direction and beyond  a critical velocity, the system can acquire a proper $S$ that triggers sufficient attractive $g_{\rm eff}$ to support self-bound droplet, which cannot be achieved if $v$ moves along other directions. This demonstrates the physical origin of chiral droplet formation.

\subsection{Full phase diagram}

To obtain the full phase diagram of gaseous and droplet states, we have computed the total energy density $\epsilon$ in terms of three variables $n,\ S$ and $p_{0x}$. At each given $n$ we minimize $\epsilon$ in terms of $S$ and $p_{0x}$, and then plot out $\epsilon/n$ as a function of $n$. A stable droplet state corresponds to a minimum of $\epsilon/n$ at finite $n$, i.e., $\partial (\epsilon/n) / \partial n=0$, which is equivalent to zero pressure ($P\equiv n\partial\epsilon/\partial n-\epsilon=0$) such that the system can be self-stabilized in vacuum. A stable gaseous state corresponds to a minimum of $\epsilon/n$ at $n\rightarrow 0$ (i.e., with $g_{\rm eff}>0$). 

\begin{figure}[t] 
    \centering 
    \includegraphics[width=9cm]{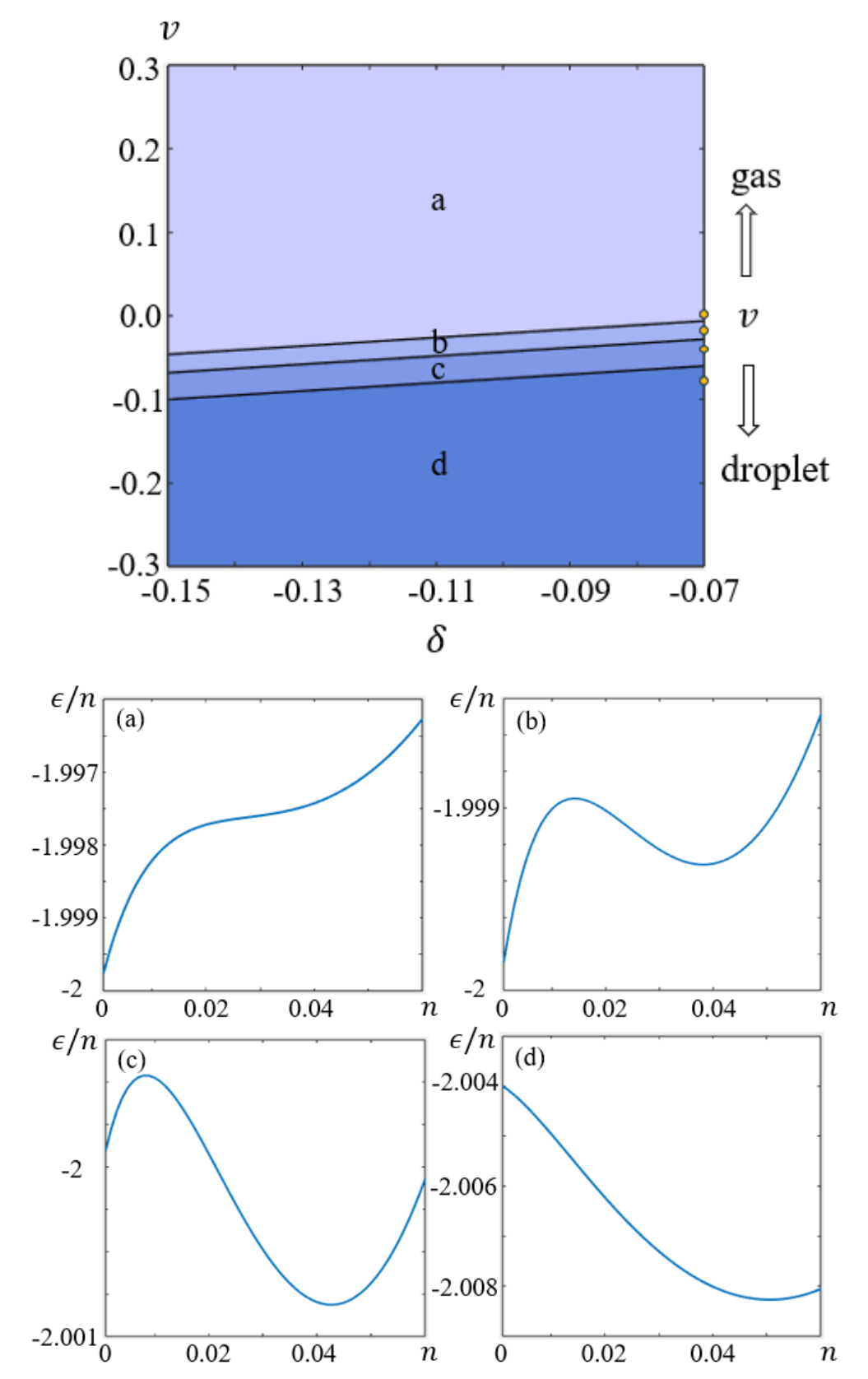} 
    \caption{(Color online.) Ground-state phase diagram in the parameter plane of magnetic detuning ($\delta$, in unit of $E_q$) and moving velocity ($v$, in unit of $q/m$). Four phases are found: (a) a single gaseous ground state, (b) a gaseous ground state and a droplet metastable state; (c) a droplet ground state and a metastable gaseous state; (d) a single droplet ground state. Their typical $\epsilon/n\sim n$ are plotted in (a,b,c,d) accordingly, at a fixed $\delta=-0.07E_q$ and different $mv/q=0 (a), -0.02 (b), -0.04(c), -0.08(d)$. Here the units of $\epsilon/n$ and $n$ are $E_q$ and $q^2/(4\pi a_B)$, respectively. 
    The other parameters are the same as  in Fig.\ref{fig_spectrum}. Different colors in the diagram are just to distinguish different phases. } \label{fig_diagram} 
\end{figure}

Fig.\ref{fig_diagram} shows the full phase diagram in the parameter plane of magnetic detuning ($\delta$) and moving velocity ($v$). Four phases are presented: a single gaseous ground state (a), a gaseous ground state and a droplet metastable state (b), a droplet ground state and a metastable gaseous state (c) and a single droplet ground state (d).  Their typical $\epsilon/n\sim n$ functional is shown in Fig.\ref{fig_diagram}(a,b,c,d) individually. Here (b) and (c) can both support liquid-gas coexistence, and the first-order liquid-gas transition occurs at the boundary between  (b) and (c). The phase boundary between (c) and (d) is equivalent to requiring  $g_{\rm eff}=0$ (see Eq.(\ref{geff})), i.e., the critical point of locally stable/unstable gaseous state.

Physically, the liquid-gas first-order transition and phase coexistence can be traced back to the ``spin twist" mechanism as pointed out in Ref.\cite{Gu}, where the single-particle potential and the interaction favor different spin polarizations and therefore the spin orientation acquires a twist as the system evolves from low to high density regime. This spin twist results in an effective repulsion in the low density regime to stabilize the gas state, so that it can coexist with droplet in high density regime  and lead to  liquid-gas coexistence. Here,  SOC plays the similar role of Rabi coupling in \cite{Gu} to generate the spin twist.

The phase diagram in Fig.\ref{fig_diagram} confirms the formation of chiral quantum droplet. Specifically, starting from a static situation ($v=0$), the gaseous state (a) will be kept if the system moving along $+x$ ($v>0$), while gradually evolve to a droplet state ($a\rightarrow b\rightarrow c\rightarrow d$) if moving along $-x$ ($v<0$). We note that a recent work\cite{Ohberg}  also reported the formation of chiral droplet in 1D by employing a model Hamiltonian with density-dependent gauge potential and interactions in both two- and three-body channels. In comparison, here we have considered a realistic system with spin-orbit coupling and with only two-body interactions, and moreover, the chiral droplet here is in 3D free space. 

 In realistic experiment, the gas and droplet phases can be well distinguished from expanding dynamics of bosons after released from traps\cite{Tarruell_Science2018,Tarruell_PRL2018,Fattori_PRL2018,Fort_PRR2019, Fort_CM2020, Wang_PRR2021}. The chiral droplet then corresponds to a stable self-bound object (without expanding) only when the system is moving towards a particular direction in free space.  On the other hand, in the presence of an external trap, the liquid-gas transition and coexistence can be detected via density discontinuity at certain location inside the trap\cite{Gu, Giorgini}, where droplet (with high density) and gas (low density) share the same chemical potential and the same pressure so that they can coexist with each other. The liquid-gas transition and coexistence can be driven either by external fields\cite{Gu} or by temperature\cite{Giorgini}, both of which produce similar $P$-$V$ isotherms as the tuning parameter changes.  

\bigskip

\section{Final remarks} \label{sec_summary}

We remark that in order to support a chiral droplet here, the interaction strengths in different spin channels have to fulfill $g_{\uparrow\uparrow},g_{\downarrow\downarrow}>0$, $g_{\uparrow\downarrow}<0$ and $g_{\uparrow\downarrow}^2>g_{\uparrow\uparrow}g_{\downarrow\downarrow}$. This is to say, the same system without moving or SOC is able to support a self-bound droplet in vacuum. We note that this requirement is not satisfied by the interaction parameters in chiral soliton experiment\cite{Frolian}, where one $g_{ii}$ is negative and the according thermodynamic system is  unstable against density fluctuations. Moreover, our calculation has been based on the full treatment of all momentum states in different helicity branches, rather than the perturbative theory around single-particle condensation momentum within the lowest helicity branch\cite{Frolian}. These differences well distinguish the two chiral phases reported in our work and the earlier experiment\cite{Frolian}.

Finally, although in this work we have particularly focused on the chiral droplet formed in SOC systems, we expect such chiral phenomenon universally applied to  quantum systems with broken Galilean invariance. For these systems, moving velocity can be used as a convenient tool to change the internal structure of the system, which, in combination with interactions, can engineer exotic chiral states. In this way, our current work opens up new opportunities for exploring intriguing topological phases under a well-controlled motion. For instance, when combine the current SOC with a field gradient along another direction, a synthetic magnetic field can be generated\cite{SMF_expt}. In this case, the system carries a topological charge and the Galilean invariance is broken in two directions. One can then expect even rich chiral phases therein due to the interplay of moving velocity and topological character, which deserve a separate study in future.

Data that support the findings of this article are openly available\cite{data}.

\acknowledgments
We thank Qi Gu for helpful discussions. This work is supported by National Natural Science Foundation of China (92476104, 12134015) and Innovation Program for Quantum Science and Technology (2024ZD0300600).

\end{document}